# Inverse Design of an All-Dielectric Nonlinear Polaritonic Metasurface

Simon Stich,[1, +] Jewel Mohajan,[2, +] Domenico de Ceglia,[3] Luca Carletti,[3] Hyunseung Jung,[4] Nicholas Karl,[5] Igal Brener [4, 5], Alejandro W. Rodriguez, [2, *] Mikhail A. Belkin,[1, *] and Raktim Sarma[4, 5,*]

[+]Equal Contribution

*E-mail addresses: rsarma@sandia.gov, mikhail.belkin@wsi.tum.de, arod@princeton.edu

[1]Walter Schottky Institut, Technische Universitat Munchen, Garching, Bavaria, Germany

[2]Department of Electrical and Computer Engineering, Princeton University, Princeton, New Jersey, USA

[3]National Institute of Optics - National Research Council (INO-CNR) and Department of Information Engineering, University of Brescia, Brescia, Italy

[4]Center for Integrated Nanotechnologies, Sandia National Laboratories, Albuquerque, New Mexico, USA

[5]Sandia National Laboratories, Albuquerque, New Mexico, USA

# Inverse Design of an All-Dielectric Nonlinear Polaritonic Metasurface

## Abstract

Nonlinear metasurfaces offer a new paradigm to realize optical nonlinear devices with new and unparalleled behavior compared to nonlinear crystals, due to the interplay between photonic resonances and materials properties. The complicated interdependency between efficiency and emission directionality of the nonlinear optical signal on the existence, localization, and lifetimes of photonic resonances, as well as on the nonlinear susceptibility, makes it extremely difficult to design optimal metasurfaces using conventional materials and geometries. Inverse design using topology optimization is a powerful design tool for photonic structures, but traditional approaches developed for linear photonics are not suitable for such high dimensional nonlinear problems. Here, we use a topology optimization approach to inverse-design a fabrication-robust nonlinear metasurface that includes quantum-engineered resonant nonlinearities in semiconductor heterostructures for efficient and directional second harmonic generation. Furthermore, we also demonstrate that under practical constraints, among all the parameters, the nonlinear modal overlap emerges as the dominant parameter that enhances conversion efficiency, a finding that contrasts with intuition-driven studies that often emphasize Purcell enhancement. Our results open new opportunities for optimizing nonlinear processes in nanophotonic structures for novel light sources, quantum information applications, and communication.

## Introduction

Second-harmonic generation (SHG) is of paramount importance for a plethora of applications ranging from nonlinear spectroscopy [1], biosensing [2], imaging [3], to pulse characterization [4]. Conventionally, given the inherently weak nonlinear response of conventional nonlinear materials, high pump light intensities incident on optically thick bulk crystals and complex phase-matching techniques are typically required to generate second harmonic (SH) signals efficiently. All-dielectric metasurfaces have recently led to a paradigm shift in nonlinear optics, as these metasurfaces can support different types of localized and de-localized photonic resonances capable of dramatically enhancing light-matter interactions [5-7], enabling significant nonlinear responses over much smaller optical volumes compared to bulk crystals while relaxing cumbersome phase matching constraints [8]. In addition, compared to their plasmonic counterparts, all-dielectric metasurfaces exhibit much lower loss and higher damage thresholds which are crucial for nonlinear and quantum optics applications which demand high pump intensities [8-12].

For any nonlinear frequency conversion process in an all-dielectric metasurface, the two most crucial quantities of interest for technologically relevant applications are the nonlinear generation efficiency and the emission directionality of the generated signal. For SHG, these quantities are determined by the photonic resonances supported by the metasurface and the tensor components of the second order nonlinear susceptibility $\chi^{(2)}$. Specifically, the SHG efficiency in an all-dielectric metasurface is proportional to the product $|(\chi^{(2)})|^2 L_F^4 L_{SH}^2 |\beta|^2$, where $L_F$ and $L_{SH}$ are the associated field enhancements at the fundamental and second harmonic frequency and $\beta$ is a nonlinear coupling factor that depends on a spatial nonlinear overlap-integral between the two modes [13]. Although the unit cell spacing in metasurfaces is usually such that pump wavelength does not experience diffraction, diffraction often occurs for the nonlinearly generated SH waves. The SH signal is therefore often emitted as multiple diffracted beams with the dominant diffractive order(s) being determined by the nonlinear crystal orientation and spatial symmetry of the resonators constituting the metasurface.

Given the large number of parameters and figures of merit involved in this process, past intuition-based approaches often fail to address all relevant criteria and thus likely lead to highly suboptimal structures. Furthermore, in addition to optimizing across all the parameters to balance the requirements of high conversion efficiency and tailored emission directionality, the design of the nonlinear metasurface should also meet practical implementation constraints. However, determining the relative importance of each parameter for achieving a design that fulfils all these criteria is difficult, making intuition-based design approaches inadequate. For these reasons, to date, all experimental demonstrations of SHG using all-dielectric metasurfaces have concentrated on enhancing either efficiency or emission directionality, but not both. In addition, these approaches have mostly focused on enhancing only one of the parameters at a time that contribute to the SHG efficiency or emission directionality. For example, recently bound-state-in-the-continuum (BIC) photonic modes have been utilized to achieve high quality factors at the fundamental wavelength [14-17]. Similarly, there has also been work on optimizing the nonlinear overlap-integral or fabricating multilayer metasurfaces [18, 19]. Other approaches have instead sought to enhance the intrinsic $\chi^{(2)}$ of materials systems, such as by using two-dimensional transition-metal-dichalcogenides as the nonlinear medium [20].

Recently, we utilized quantum-engineered resonant nonlinearities supported by intersubband transitions (ISTs) in semiconductor heterostructures to boost $\chi^{(2)}$ by orders of magnitude and demonstrated record-high SHG efficiency using all-dielectric metasurfaces [21-23]. However, in these works [21-23], emission directionality could not be optimized leading to negligible SH emission along the normal direction when the metasurface was pumped at normal incidence. While control of emission directionality of SH signals has been explored using different materials [24], crystal orientations of III-V semiconductors [25], oblique-incident pump angles, and resonator designs [26-29], the total SHG efficiency of dielectric metasurfaces was not shown to be significantly enhanced.

In this work, we use a fundamentally new methodology for inverse-designing all-dielectric nonlinear metasurfaces for optimizing both SHG efficiency and directionality and then we experimentally implement and characterize the optimized metasurface designs. The computational approaches for discovering optical structures based on desired functional characteristics have recently redefined the limits of nanophotonic devices in the linear regime [30, 31]. While there also have been theoretical investigations of these approaches for optimizing nonlinear optical devices, including metasurfaces [32, 33], to date, there has been no inverse design approach for nonlinear optics that considers implementation constraints for practical applications followed by an experimental demonstration. Here, we demonstrate a metasurface capable of achieving large field enhancements at the fundamental and SH frequency, strong nonlinear overlaps between the two modes, and high emission directionality, while simultaneously controlling for the thickness and periodicity of the metasurface so as to be as small as possible. Our inverse design approach consists of two sequential steps, where the first step focuses on enhancing the SHG efficiency while taking into account the fabrication constraints followed by the second step that focuses on optimizing the emission directionality. Since our approach considers fabrication tolerances, the predicted nanophotonic structure is both efficient and experimentally realizable, enabling record-high SHG efficiency for dielectric metasurfaces and SH emission along the normal direction. Furthermore, we demonstrate that the inverse design approach when implemented with practical constraints, also allows us to have a deeper understanding of the importance of different parameters that dictate the light-matter interaction for nonlinear optics. *Specifically, we show that under practical implementation constraints, among all the parameters, the nonlinear modal overlap emerges as the dominant parameter that enhances conversion efficiency, a finding that contrasts with intuition-driven studies that often emphasize Purcell enhancement or high-quality factors of the optical modes.* Our results therefore open exciting new directions for developing classical or quantum light sources based on nonlinear processes in dielectric metasurfaces such as harmonic generation, mixing or spontaneous parametric down conversion [10-12]. For example, photon-pair sources based on nonlinear-metasurfaces with higher photon rates could be designed using the formalism presented in this paper [10-12]. Such light sources can be inverse-designed to be in the mid to long-infrared region of the electromagnetic spectrum where currently there is a lack of compact and efficient light sources for numerous applications ranging from sensing [34], non-destructive testing [35] to optical coherence tomography [36].

## Inverse Design of the Nonlinear Polaritonic Metasurface

The net nonlinear response of a metasurface is a consequence of a complicated light-matter interaction, wherein the matter component determines the elements of $\chi^{(2)}$ tensor, and the photonic component determines possible field components enhancements and optical-mode symmetry. To maximize the overall nonlinear response, we optimized both facets. For the matter component, we designed and optimized quantum engineered ISTs in multi-quantum-well (multi-QW) semiconductor heterostructures for generating a giant resonant nonlinear response. ISTs offer one of the largest $\chi^{(2)}$ available in condensed matter systems [37], and they have been recently utilized in light-matter coupled plasmonic and all-dielectric metasurfaces to achieve record-high second-order [21-23, 38-40] and third-order nonlinear responses [41-43].

Figure 1(a) shows a schematic of the light-matter coupled polaritonic metasurface optimized for enhancing SHG efficiency and emission directionality. The individual resonators of the metasurface are made out of a nonlinear medium comprised of a multi-QW heterostructure which was optimized so as to maximize the magnitude of $\chi^{(2)}$ at specific wavelengths. The metasurface is fabricated on a sapphire substrate which provides the refractive index contrast with the multi-QW heterostructure needed for the resonators to support a photonic mode, along with a $SiO_2$ capping layer which we utilize as a mask for fabrication. We design the multi-QW heterostructure to support three equally spaced electronic energy levels with energy separation of ~ 156 meV (7.9 µm). Our coupled QW design is shown in Fig. 1(b). It lacks inversion symmetry along the growth direction *z*, which allows us to have optical transitions at both the fundamental pump wavelength (7.9 µm) and SH wavelength (3.95 µm) from the ground state and thereby creates a doubly-resonant $\chi^{(2)}$ [22]. The IST selection rules restrict the polarization of the pump light that can couple to these electronic transitions to z direction. Therefore the only significant non-zero element of the rank-3 $\chi^{(2)}$ tensor in our multi-QW heterostructure is $\chi^{(2)}_{zzz}$ and is given by the expression:

$$\chi^{(2)}_{zzz} = \frac{e^3}{\epsilon_0 \hbar^2} \frac{N_d z_{12} z_{23} z_{31}}{(\omega - \omega_{12} - 2i\gamma_{12})(2\omega - \omega_{13} - 2i\gamma_{13})} \quad (1)$$

where $e$ is the elementary charge, $\epsilon_0$ the vacuum permittivity, $\hbar$ the reduced Planck constant, $N_d$ the doping density, $2\gamma_{ij}$ the IST damping rate due to the resonant absorption, and $z_{ij}$ the dipole matrix lengths between the level $i$ and $j$. Figure 1(b) shows an 8-band k.p band structure calculation of a single period of the multi-QW system used in this study and the corresponding $\chi^{(2)}_{zzz}$ calculated using Eq. (1), which can be as large as ~ 200 nm/V, is shown in Fig. 1(c).

While having an IST-based nonlinear medium satisfies the criteria of having a giant $|\chi^{(2)}|$, it has well-known associated drawbacks which we need to overcome to realize an all-dielectric metasurface with optimum SHG efficiency and emission directionality. In particular, ISTs only have large transition moments polarized in the z-direction, which makes it difficult to couple them to normally-incident light. Furthermore, resonant IST nonlinearities produce absorption at pump and SH frequencies which reduces the field enhancement of the photonic resonances [22]. For optimum SHG efficiency, the resonators should support photonic resonance with spatial field profiles that maximize the product $(|\chi^{(2)}_{zzz}|)^2 L_F^4 L_{SH}^2 |\beta|^2$. $L_F$ and $L_{SH}$ in this case are the associated field enhancements of the z-component of the fields at the fundamental and second harmonic frequencies given by $\frac{|E_z^\omega|}{|E_{inc}|}$ and $\frac{|E_z^{2\omega}|}{|E_{inc}|}$, respectively ( $|E_{inc}|$ is the incident electric field), and $\beta$ is proportional to the spatial overlap integral given by $\frac{\int dV \psi (E_{SH}^{2\omega})^* (E_F^\omega)^2}{(\int dV \epsilon_F |E_F^\omega|^2)(\sqrt{\int dV \epsilon_{SH} |E_{SH}^{2\omega}|^2})}$, where $E_F^\omega$ and $E_{SH}^{2\omega}$ are the fields at the fundamental and second harmonic respectively, $\psi = 1$ inside the nonlinear medium and zero elsewhere, and $\epsilon_F$ and $\epsilon_{SH}$ are permittivities of the multi-QW heterostructure at the fundamental and second harmonic frequencies, respectively [13].

Since the only non-zero element of the $\chi^{(2)}$ tensor is $\chi^{(2)}_{zzz}$, for normally incident pump and conventional symmetrical resonator geometries such as cylinders, symmetry prohibits the emission of SH signal along the normal direction (i.e. along the zero-diffraction order) which is highly desirable for applications. This implies, for optimum SHG efficiency and normal emission, the photonic resonators must lack inversion symmetry while possessing resonances at both pump and SH frequency with large overlap integral $\beta$. Since optimization of such resonator designs is extremely complex with several interdependent parameters, we resort to a computational inverse design approach.

The core idea behind our inverse design approach is to formulate the metasurface unit cell resonator design process as an optimization problem where the design goal is first expressed as a function of the resonator fields and then maximized by tuning the geometry of the photonic resonator structure. While there exist many inverse design methods, in this work, we utilize topology optimization (TO) [30, 31], which has recently been demonstrated to be extremely effective in producing optimum designs with performances close to the possible theoretical limit [44- 46]. Within TO, the whole computational domain of the problem is spatially discretized using high enough resolution so that the fields are properly resolved, and the density of the material at each voxel in the design domain is considered as the degree of freedom, $\rho \in [0, 1]$ where 0 means absence of the multi-QW heterostructure material or free space and 1 means presence of the multi-QW heterostructure. Figure 1(d) shows a schematic of the optimization process, where the optimization selects for the presence or absence of material at every voxel of the discretization grid, which in our case was chosen to be a 50 nm square grid along both *x* and *y* dimensions. Note that unlike optimization of SHG in plasmonic metasurfaces [32], our all-dielectric metasurface requires full three-dimensional optimization because fields are confined throughout the entire metasurface as opposed to being evanescent.

Once the design domain of the problem is specified, Maxwell's equations are solved in the frequency domain for a planewave source (incident normally onto the metasurface from the top) to calculate field $E_F^\omega$ at the fundamental frequency $\omega_F$ (corresponding to pump wavelength = 7.9 µm). All calculations are performed under the assumptions of non-depleted pump [44]. To incorporate the degrees of freedom $\rho$, the electric permittivity in the computation region is defined as $\epsilon(x, y, z) = \epsilon_0 + \rho_{(x,y,z)}\left(\epsilon_{(x,y,z)}^{QW} - \epsilon_0\right)$, where $\epsilon_0$ and $\epsilon_{(x,y,z)}^{QW}$ are permittivities of vacuum and the multi-QW respectively as shown in Fig. 1(d). The calculated value of the *z*-component of $E_F^\omega$ is then used to calculate the value of the polarization current $J_{SH} = i\omega_{SH}\epsilon_0\chi^{(2)}(E_F^\omega)^2$ at the SH frequency $\omega_{SH}$ which acts as a source for generating the electric fields $E_{SH}^{2\omega}$ at the second harmonic (see Supplementary Information for further details). Since we are interested in simultaneously optimizing the SHG efficiency and emission directionality, we carefully designed our optimization procedure and performed it sequentially in two steps. In the first step, we initialize the entire computation region with $\rho = 0.5$ and use the total power that includes both radiated and non-radiated power at the SH frequency to optimize. The total power at SH frequency is computed as $\int J_{SH}^* \cdot E_{SH}^{2\omega}\, dV$, where $J_{SH}^*$ is the *z*-component of the complex conjugate of $J_{SH}$ and the integration runs over the entire optical volume. This step allows us to have a geometry which supports modes at the fundamental frequency and SH frequency with large field enhancements and large nonlinear field overlap factors. To simultaneously ensure high SH emission intensity in vertical direction for a normal incident pump, we use the optimized metasurface unit cell design parameters from the first step as the initial unit cell design parameters for the second round of optimization. The far-field *z*-component of the SH Poynting vector flux is used as the objective to optimize for this step. For benchmarking, we compare the performance of our optimized structure with a metasurface of the same unit cell periodicity but comprising cylindrical resonators supporting a magnetic dipole-type Mie mode at the pump wavelength and made of the same IST-based nonlinear medium which recently demonstrated record high SHG efficiency [22].

The TO involves an extremely high-dimensional space, and to efficiently navigate this design space, the only viable approach is to use gradient-based optimization. Fortunately, the adjoint method allows us to compute the objective gradient at a similar cost to computing the objective function itself, through the solving of an adjoint partial differential equation (PDE) [47]. At each step of the TO, we therefore deploy the adjoint method to calculate the gradient of the objective with respect to the degrees of freedom $\rho$. The globally convergent method-of-moving-asymptotes (MMA) algorithm for gradient-based local optimization is used to find the optimum. In addition, in the density TO approach, since we allow $\rho$ to vary continuously between 0 and 1 during the optimization, the permittivity can vary continuously to make use of gradients. This can result in intermediate 'grey' structures which are not realizable. Therefore, once we reach the optimum, we use filters to smooth-out and binarize the design. The binarization filters are wrapped around the optimization process so that the structure is optimized and binarized on the fly. This ensures that we do not reduce the SHG efficiency while binarizing the design.

Finally, we point out that TO can explore a very rich design space (literally every possible structure representable on the underlying grid) without assumptions on the specific working principle of the final design. As described further below, while such an approach can predict structures which are close to being global optima [44-46], during the optimization process, we apply filters to enforce constraints incorporating fabrication limitations such as the minimum feature size, slanting, aspect ratios etc. While such filtering degrades the performance slightly, as shown later, this allows us to have optimal structures that are experimentally realizable with performances that are robust to fabrication defects and limitations. A detailed description of the TO algorithm utilized for this work is given in the Supplementary Information.

## Numerical Simulations and Experimental Results

To validate the results of the TO algorithm as well as to quantify the performance of the optimum metasurface, we first performed linear and nonlinear full-wave simulations of the benchmark metasurface with cylindrical Mie-resonators shown in Figure 2(a) and the optimized metasurface predicted by TO shown in Figure 2(f). The period of the metasurface is 4.2 µm and the thickness of the heterostructure is 1.6 µm. As shown in Figure 2, while the cylindrical resonators support a magnetic dipole-type Mie mode at the pump frequency [22], the TO structure supports modes resulting in significantly larger field enhancements for the out-of-plane polarized fields at both the pump (Fig. 2(b,g) and Fig. 2 (d,i)) and second harmonic (Fig. 2(c,h) and Fig. 2(e,j)) frequencies within the entire periodic volume. In addition, in Figure 3, we plot and compare the calculated SH power spectra emitted by the two metasurfaces in reflection along the different diffraction orders. As shown in Figure 3(a), while the metasurface with cylindrical resonators has no SH emission along the normal direction, the inverse-designed metasurface emits a significant portion (~ 10 %) of the total SH power along the normal direction. In fact, the zero-diffraction order for the inverse-designed metasurface has comparable efficiency to the first diffraction orders shown in Figures 3(b-e). In addition, as shown in Figures 3(b-e), all diffraction orders of the TO designed metasurface emit larger SH power compared to the cylindrical metasurface, resulting in overall larger SHG efficiency. Figure 3(f) plots and compares the total SH power generated towards the pump direction for the metasurface with cylindrical and inverse-designed structures. At the optimized wavelength ($\lambda_{IST}$ = 7.9 µm), the inverse-designed structure is ~ 11x more efficient compared to the cylinders. Note that, due to Rabi splitting resulting from strong light-matter coupling, the cylindrical metasurface has a dip at 7.9 µm in the SH spectrum and the peak of the SH power is at a slightly shorter wavelength at $\lambda_C$. Compared to the maximum SHG efficiency of the cylindrical metasurface at $\lambda_C$, the inverse-designed structure has ~ 5x higher SHG efficiency.

Next, we performed experiments to validate our predictions. The fabrication details of the experimentally realized metasurfaces are given in Methods. First, for benchmarking, we fabricated metasurfaces with cylindrical resonators with three different radii such that the photonic resonance spans the entire spectral linewidth of the IST resonance. This allowed us to select the most efficient metasurface with cylindrical resonators and we used those results as our benchmark to compare to the performance of the inverse-designed metasurface. The details of the measurements of the metasurfaces with cylindrical resonators are given in Supplementary Information.

4(a, b) compare the simulated (Fig. 4(a)) and experimentally measured (Fig. 4(b)) linear reflectance spectrum of the inverse-designed metasurface. The metasurface has a mode $M_1$ that spectrally overlaps with the IST resonance which shows up as a dip in the reflection spectrum. We observe good qualitative agreement between simulations and experiments. Figure 4(c) shows the experimentally measured SH spectrum from the inverse-designed metasurface. Similarly to the metasurface with cylinders of optimum radius (see Supplementary Information section S1), the inverse-designed metasurface yields a maximum SHG efficiency at the pump wavelength $\lambda_P \sim 8.1$ µm. To compare the SHG efficiency to our benchmark, we fixed the pump wavelength at $\lambda_P$, and measured the SH power as a function of pump power for both designs. We performed this measurement using two microscope setups. In the first setup, we used a lens of numerical aperture (NA) = 0.18 to both pump and collect the SH signal. In the second setup, we used a lens with NA = 0.85 (see Methods for details). The first setup (NA = 0.18) allows us to collect SH signal emitted only in the zero-diffraction order whereas a lens with NA= 0.85 allows us to collect SH signal emitted in both the zero and first-diffraction orders. Figure 4(d) shows the experimental results with NA = 0.18 where we see that the SH power from the inverse-designed metasurface increases linearly with the square of the pump power. More importantly, in this experimental configuration, we observe that the inverse-designed metasurface shows two orders of magnitude larger SH signal compared to that of the metasurface comprising cylindrical resonators. Finally, we repeated the SH measurement using the lens with NA = 0.85 and the results are shown in Figure 4(e). The inverse-designed metasurfaces now show approximately 2.5 times higher SHG efficiency compared to the metasurface made of cylindrical resonators with normalized SHG efficiency at low pumping intensities of 0.38 mW/W$^2$ vs 0.15 mW/W$^2$. Notably, the inverse-design metasurface reported in Fig. 4 shows the highest SHG efficiency reported for all-dielectric nonlinear metasurfaces to date. In addition, the SHG efficiency of the inverse-design metasurface for the zero-order emission is 0.047 mW/W$^2$, which as predicted by simulations, is ~ 10% of the total SHG efficiency. We note that the simulations shown in Fig. 3 predicted approximately 5 times larger SHG efficiency for the inverse-designed structure compared to the experimentally measured results. We attribute this discrepancy to the fabrication imperfections. Figure 5(f) also shows that SHG conversion efficiency saturates at higher pump powers; this effect has previously been observed in plasmonic intersubband metasurfaces and explained by the intensity saturation of the ISTs [48]. For measurements with NA=0.85, because of the smaller pump spot size at the focal plane (and therefore larger intensities at the focus for same pump power), compared to the measurement with NA = 0.18, the measured SH power increases sub-linearly with the square of the pump power. This effect is expected to be more dominant for the inverse-designed metasurface because of the presence of modes supporting larger field enhancements.

## Importance of Constraints on Inverse Design

In the previous section, we intentionally imposed stringent fabrication constraints on the TO algorithm. As a concluding part of this work, we demonstrate the importance of introducing these constraints in practical applications. Figure 5(a) shows the optimum structure predicted by TO in the absence of such fabrication constraints. Compared to the previous inverse-designed structure (Figure 2(f)), the new structure looks more complicated, is non-contiguous, and has numerous isolated empty regions. We confirmed via simulations that the non-contiguous spatial profile and isolated empty regions are indeed required to achieve high SHG efficiency. In fact, we observed in simulations that removing some of these isolated regions of the optimized structure significantly reduces the SHG efficiency to values below the SHG efficiency of the structure shown in Figure 2(f). This implies introduction of constraints during the optimization, as opposed to imposing constraints post optimization, is crucial for achieving optimum structures. Furthermore, since the SHG efficiency depends on number of parameters, introduction of constraints in the TO algorithm changes which dominant parameter TO chooses to optimize to increase the SHG efficiency. For example, when fabrication constraints are introduced (Figure 2(f)), while all the parameters that contribute to SHG were enhanced compared to cylinders, the dominant factor that TO optimized to increase SHG is $\beta$, which for the structure shown in Figure 2(f) is approximately 40 times larger compared to the metasurfaces with cylinders (Figure 2(a)). As opposed to that, for the optimization with no constraints, the dominant factor that TO chooses to optimize is the quality factor ($Q$) of the mode at the fundamental wavelength which for the structure shown in Fig. 5(a) is approximately 9 times larger compared to cylinders ($\beta$ is enhanced only by a factor of 2.6 in this case). Since photonic modes with a high $Q$ factor are more sensitive to fabrication defects, this makes the un-constrained inverse-designed structure significantly more difficult to be experimentally realized. We confirm this by fabricating and characterizing the metasurface shown in Fig. 5(a).

Figure 5(b) shows the calculated linear reflection spectrum of the structure, and we see that the photonic structure supports modes $M_1$, $M_2$, $M_3$, and $M_4$ within the spectral window of 6-9 µm, and the mode $M_1$ spectrally overlaps with the IST resonance for efficient SHG. Figure 5(c) plots and compares the calculated total reflected SH power for the metasurface with cylindrical and the two inverse-designed metasurfaces. As expected, because of lack of constraints, at the optimized wavelength (7.9 µm), the inverse-designed structure designed without constraints is approximately 17 times more efficient compared to the cylinders. Compared to the peak of SHG at $\lambda_C$ for the cylindrical metasurface, the same inverse-designed structure has approximately 7.5 times higher SHG efficiency compared to cylinders.

While the unconstrained inverse-designed structure is significantly more efficient in theory, its complicated spatial profile makes it difficult to realize experimentally. In particular, the isolated small empty regions demand aspect ratios that are approximately 1:16 which are extremely hard to achieve in fabrication. Figure 5(d) shows SEM image of the inverse-designed metasurface which we fabricated using our most optimum fabrication and etching recipe. While the fabricated metasurface unit cell shown Fig. 5(d) overall resembles the target spatial profile shown in Fig 5(a), the fabrication cannot completely resolve all the small isolated empty regions. Consequently, this leads to having more nonlinear material in the inverse-designed structure than required and leads to shifting of the modes to a longer wavelength. Figure 5(e) shows the experimentally measured linear spectra of the metasurface. While we observe the presence of all the four modes ($M_1$, $M_2$, $M_3$, $M_4$) within the spectral window of 6-9 µm, because of the red shift, now the mode $M_2$ instead of $M_1$ spectrally overlaps with the IST resonance. Since the structure is optimized for $M_1$, this dramatically reduces the SHG efficiency and the experimentally measured SHG efficiency is lower than even cylinders as shown in Fig 5 (f). We note that in principle, if the exact limitations of fabrication are known (e.g. etching profile, blurring etc.) it may be possible to pre-compensate for the red shifting of the modes because of fabrication imperfections in the TO algorithm as perturbative effects. However, given the large number of parameters involved, it will require a much more elaborate computational framework than the presented framework and will be investigated in our future studies.

## Conclusion

To conclude, we have demonstrated an experimentally realizable inverse design approach for an all-dielectric polaritonic nonlinear metasurface that utilizes quantum-engineered intersubband transitions in semiconductor heterostructures to produce a giant resonant nonlinearity. Along with maximizing the nonlinear generation efficiency, we utilize a two-step sequential topology optimization technique to simultaneously optimize the directionality and the efficiency of the SH emission. Our design is robust to fabrication defects, shows a record-high SHG efficiency for all-dielectric metasurfaces of 0.38 mW/W$^2$, and generates a significant portion of the SH power along the normal direction for normally-incident pump, which is otherwise forbidden by symmetry of the nonlinear susceptibility. Our inverse design approach, in addition to realizing nonlinear devices with record-high efficiencies and novel functionalities such as directional emission, also demonstrate that under practical implementation constraints, among all the parameters, the nonlinear modal overlap emerges as the dominant parameter that enhances conversion efficiency, a finding that contrasts with intuition-driven studies that often emphasize Purcell enhancement or high-quality factors of the optical modes. Finally, we have shown that, depending on what fabrication constraints we impose on the TO algorithm, it may be possible to realize designs that may have larger SHG efficiencies suggesting further room for improvement. Nevertheless, the emission directionality and efficiency of the inverse-designed metasurface presented in this work are already approaching metrics that are highly sought after for various applications. Our results therefore open exciting new opportunities for optimizing nonlinear and quantum processes such as spontaneous parametric down conversion in metasurfaces using TO design algorithms for multiple applications ranging from biosensing, spectroscopy to quantum information.

## Methods

**Full-Wave Numerical Simulations**

The full wave linear and nonlinear numerical simulations were performed using the frequency domain, finite-element method (COMSOL). The multi-QW embedded inside the resonators was modeled as a homogeneous, anisotropic layer with dielectric-constant tensor $\epsilon_{(x,y,z)}^{QW} = \epsilon_t(\hat{x}\hat{x} + \hat{y}\hat{y}) + \epsilon_l\hat{z}\hat{z}$, where the transverse dielectric constant is $\epsilon_t = 11.08$ and the longitudinal dielectric constant is

$$\epsilon_l = 11.08 + \frac{f_{12}\omega_{12}^2}{\omega_{12}^2-\omega^2-2i\omega\gamma_{12}} + \frac{f_{13}\omega_{13}^2}{\omega_{13}^2-\omega^2-2i\omega\gamma_{13}}.$$

In the expression above, the first Lorentzian term models the first optical transition at 7.9 μm ($\omega_{12}$ =2π×38.4×10$^{12}$ rad/s) and the second term models the transition at SH wavelength at 3.95 μm ($\omega_{13}$ =2π×76.8×10$^{12}$ rad/s). $f_{12}$ and $f_{13}$ are the oscillator strengths and are proportional to the product of the doping density of the quantum wells and the dipole matrix length for the corresponding transition. The term $2\gamma_{12}$ and $2\gamma_{13}$ represent the IST damping rate due to the resonant absorption and were determined by experimental measurements. The dipole matrix lengths, IST resonant frequencies, and band structures shown in Fig. 1(b) were determined from calculations using NextNano3, a commercial software by nextnano GmbH. The dipole matrix lengths between the level $i$ and level $j$ for the optimized QW structure used in this study are $z_{12} = 1.6$ nm, $z_{23} = 1.99$ nm and $z_{31} = 0.92$ nm. For the nonlinear simulations, the resonant, quadratic nonlinear susceptibility of the multi-QW, $\chi_{QW}^{(2)} = \chi^{(2)}\hat{z}\hat{z}\hat{z}$, was modeled based on Equation 1. The nonlinear simulations were done in the undepleted pump approximation and neglecting saturation effects of the ISTs. For simulating the linear fields at the fundamental frequency, the electromagnetic problem was solved in a unit cell using Floquet boundary conditions on the planes $x = 0$, $y = 0$ and $x = p$, $y = p$, where $p$ is the periodicity equal to 4.2 μm. Port boundary conditions were used to launch the input pump as a plane wave polarized along x-direction at normal incidence to calculate the linear reflectance spectrum. For calculating the field at the second harmonic frequency, the solution at the pump was then used to calculate the second-harmonic, nonlinear current source for the electromagnetic problem at the second-harmonic frequency. At the second-harmonic, Floquet boundary conditions were used on the $x = 0$, $y = 0$, $x = p$ and $y = p$, planes. Plane-wave ports adapted to each diffraction order were used on the air side and sapphire side.

The SH emitted power along each diffraction order for the cylinders and the inverse-designed structures shown in the main text were calculated by integrating the Poynting vector associated with the corresponding plane-wave port. The overall efficiency in is calculated as the sum of all the diffraction order efficiencies (either on the air side or on the sapphire side).

**Fabrication of the Nonlinear Metasurfaces**

The metasurfaces were fabricated by epi-side down adhesive bonding with Bencocylcobutene (BCB) to sapphire and a subsequent substrate removal process consisting of lapping and selective chemical wet etching. The resulting structure consists of a 1.6μm MQW thin film on sapphire with an intermediate layer of BCB. The inverse design pattern is transferred onto the MQW by electron-beam lithography with hydrogen silsesquioxane (HSQ) followed by a reactive ion etching (RIE) process using Chlorine and Argon as etchants.

**Nonlinear Optical Measurements**

To evaluate the second harmonic performance, the metasurface was pumped using a tightly focused pulsed tunable QCL (Daylight Solutions, Inc.; tuning range 3-13μm, pulse length 400 ns, and repetition rate 250 kHz). The generated nonlinear radiation is reflected into a power-calibrated liquid $N_2$-cooled MCT detector with a beam splitter, that separates the pump beam from the nonlinear beam. Any residual pump radiation is blocked by a suitable filter in front of the detector. The pump beam was focused onto the metasurface using an aspheric lens with two different numerical apertures (NA) = 0.18 and 0.85 respectively. Since the aspheric lens acts both as a focusing lens for the pump radiation and as a collecting lens for reflection SH radiation, the detected diffraction orders can be controlled by varying the NA. The anti-reflection coating was chosen for the SH wavelength to ensure the maximum collection efficiency of the SH signal. For the lens with NA= 0.85, the tightly focused spot size at the sample position was determined experimentally with the knife-edge technique to be $2w = 36 \,\mu m$. The spot size is inversely proportional to the NA of the used focusing lens and is 4.7 times larger for the NA = 0.18 lens. Thus, the pumping intensity at the sample position is 22 times lower for the lower NA lens compared to the high NA lens.

# References

[1] Wang, Y., Xiao, J., Yang, S., Wang, Y., Zhang, X. Second harmonic generation spectroscopy on two-dimensional materials. *Optical Materials Express* **9 (3),** 1136-1149 (2019).

[2] Tran, R. J., Sly, K. L., & Conboy, J. C. Applications of surface second harmonic generation in biological sensing. *Annual Review of Analytical Chemistry*, *10*(1), 387-414 (2017)/

[3] Aghigh, A., Bancelin, S., Rivard, M., Pinsard, M., Ibrahim, H., & Légaré, F. Second harmonic generation microscopy: a powerful tool for bio-imaging. *Biophysical Reviews*, *15*(1), 43-70 (2023).

[4] Li, Y., Chen, Y., Li, W., Wang, P., Shao, B., Peng, Y., & Leng, Y. Accurate characterization of mid-infrared ultrashort pulse based on second-harmonic-generation frequency-resolved optical gating. *Optics & Laser Technology*, *120*, 105671 (2019).

[5] Brener, I., Liu, S., Staude, I., Valentine, J., Holloway, C. Dielectric Metamaterials: Fundamentals, Designs and Applications (Woodhead Publishing, 2019)

[6] Kuznetsov, A. I., Brongersma, M. L., Yao, J., Chen, M. K., Levy, U., Tsai, D. P., ... & Pala, R. A. Roadmap for optical metasurfaces. *ACS photonics*, *11*(3), 816-865 (2024).

[7] Schulz, S. A., Oulton, R., Kenney, M., Alù, A., Staude, I., Bashiri, A., ... & Fernandez-Corbaton, I. Roadmap on photonic metasurfaces. *Applied Physics Letters*, *124*(26) (2024).

[8] Krasnok, A., Tymchenko, M., Alu, A. Nonlinear Metasurfaces: A Paradigm Shift in Nonlinear Optics. *Materials Today* **21(1)**, 8-21 (2018).

[9] Tokman, M. et al. Purcell Enhancement of the Parametric Down-Conversion in Two-Dimensional Nonlinear Materials. *APL. Photonics.* **4,** 034403(2019).

[10] Santiago-Cruz, T., Fedotova, A., Sultanov, V., Weissflog, M. A., Arslan, D., Younesi, M., ... & Chekhova, M. Photon pairs from resonant metasurfaces. *Nano letters*, *21*(10), 4423-4429 (2021).

[11] Zhang, J., Ma, J., Parry, M., Cai, M., Camacho-Morales, R., Xu, L., ... & Sukhorukov, A. A. Spatially entangled photon pairs from lithium niobate nonlocal metasurfaces. *Science Advances*, *8*(30), eabq4240 (2022).

[12] Santiago-Cruz, T., Gennaro, S. D., Mitrofanov, O., Addamane, S., Reno, J., Brener, I., & Chekhova, M. V. Resonant metasurfaces for generating complex quantum states. *Science*, *377*(6609), 991-995 (2022).

[13] Lin, Z., Liang, X., Lončar, M., Johnson, S. G., & Rodriguez, A. W. Cavity-enhanced second-harmonic generation via nonlinear-overlap optimization. *Optica*, *3*(3), 233-238 (2016).

[14] Vabishchevich, P. P., Liu, S., Sinclair, M. B., Keeler, G. A., Peake, G. M., & Brener, I. (2018). Enhanced second-harmonic generation using broken symmetry III–V semiconductor Fano metasurfaces. *Acs Photonics*, *5*(5), 1685-1690 (2018).

[15] Anthur, A. P., Zhang, H., Paniagua-Dominguez, R., Kalashnikov, D. A., Ha, S. T., Maß, T. W., ... & Krivitsky, L. Continuous wave second harmonic generation enabled by quasi-bound-states in the continuum on gallium phosphide metasurfaces. *Nano Letters*, *20*(12), 8745-8751 (2020).

[16] Koshelev, K., Kruk, S., Melik-Gaykazyan, E., Choi, J. H., Bogdanov, A., Park, H. G., & Kivshar, Y. Subwavelength dielectric resonators for nonlinear nanophotonics. *Science*, *367*(6475), 288-292 (2020).

[17] Liu, Z., Xu, Y., Lin, Y., Xiang, J., Feng, T., Cao, Q., ... & Liu, J. High-Q quasibound states in the continuum for nonlinear metasurfaces. *Physical review letters*, *123*(25), 253901 (2019).

[18] Marino, G., Rocco, D., Gigli, C., Beaudoin, G., Pantzas, K., Suffit, S., ... & De Angelis, C. Harmonic generation with multi-layer dielectric metasurfaces. *Nanophotonics*, *10*(7), 1837-1843 (2021).

[19] Carletti, L., Locatelli, A., Stepanenko, O., Leo, G., & De Angelis, C. Enhanced second-harmonic generation from magnetic resonance in AlGaAs nanoantennas. *Optics express*, *23*(20), 26544-26550 (2015).

[20] Malard, L. M., Alencar, T. V., Barboza, A. P. M., Mak, K. F., & De Paula, A. M. Observation of intense second harmonic generation from $MoS_2$ atomic crystals. *Physical Review B—Condensed Matter and Materials Physics*, *87*(20), 201401 (2013).

[21] Sarma, R., De Ceglia, D., Nookala, N., Vincenti, M. A., Campione, S., Wolf, O., ... & Brener, I. Broadband and efficient second-harmonic generation from a hybrid dielectric metasurface/semiconductor quantum-well structure. *ACS Photonics*, *6*(6), 1458-1465 (2019).

[22] Sarma, R., Xu, J., De Ceglia, D., Carletti, L., Campione, S., Klem, J., ... & Brener, I. An all-dielectric polaritonic metasurface with a giant nonlinear optical response. *Nano Letters*, *22*(3), 896-903 (2022).

[23] Sarma, R., Xu, J., Ceglia, D. D., Carletti, L., Klem, J., Belkin, M. A., & Brener, I. Control of second-harmonic generation in all-dielectric intersubband metasurfaces by controlling the polarity of χ (2). *Optics Express*, *30*(19), 34533-34544 (2022).

[24] Fedotova, A., Younesi, M., Sautter, J., Vaskin, A., Löchner, F. J., Steinert, M., ... & Setzpfandt, F. Second-harmonic generation in resonant nonlinear metasurfaces based on lithium niobate. *Nano letters*, *20*(12), 8608-8614 (2020).

[25] Sautter, J. D., Xu, L., Miroshnichenko, A. E., Lysevych, M., Volkovskaya, I., Smirnova, D. A., ... & Rahmani, M. Tailoring second-harmonic emission from (111)-GaAs nanoantennas. *Nano letters*, *19*(6), 3905-3911 (2019).

[26] Frizyuk, K. Second-harmonic generation in dielectric nanoparticles with different symmetries. *JOSA B*. **36(8)**, F32-F37 (2019).

[27] Rocco, D., Gigli, C., Carletti, L., Marino, G., Vincenti, M. A., Leo, G., & De Angelis, C. Vertical second harmonic generation in asymmetric dielectric nanoantennas. *IEEE Photonics Journal*, *12*(3), 1-7 (2020).

[28] Xu, L., Rahmani, M., Smirnova, D., Zangeneh Kamali, K., Zhang, G., Neshev, D., & Miroshnichenko, A. E. Highly-efficient longitudinal second-harmonic generation from doubly-resonant AlGaAs nanoantennas. *Photonics* 5(3), 29 (2018).

[29] Carletti, L., Locatelli, A., Neshev, D., & De Angelis, C. Shaping the radiation pattern of second-harmonic generation from AlGaAs dielectric nanoantennas. *ACS Photonics*, *3*(8), 1500-1507 (2016).

[30] Molesky, S., Lin, Z., Piggott, A. Y., Jin, W., Vucković, J., & Rodriguez, A. W. Inverse design in nanophotonics. *Nature Photonics*, *12*(11), 659-670 (2018).

[31] Jensen, J. S., & Sigmund, O. Topology optimization for nano-photonics. *Laser & Photonics Reviews*, *5*(2), 308-321 (2011).

[32] Mann, S. A., Goh, H., & Alù, A. Inverse design of nonlinear polaritonic metasurfaces for second harmonic generation. *Acs Photonics*, *10*(4), 993-1000 (2023).

[33] Li, N., Zhang, J., Neshev, D. N., & Sukhorukov, A. A. Inverse design of nonlinear metasurfaces for sum frequency generation. *Nanophotonics*, in press (2024).

[34] Jung, D., Bank, S., Lee, M. L., Wasserman, D. Next-Generation Mid-Infrared Sources. *Journal of Optics.* **19(12),** 123001(2017).

[35] Qu, Z., Jiang, P., Zhang, W. Development and Application of Infrared Thermography Non-Destructive Testing Techniques. *Sensors*. **20(14)**, 3851 (2020).

[36] Colley, C. S. et al. Mid-Infrared Optical Coherence Tomography. *Review of Scientific Instruments*, **78(12)**, 123108 (2007).

[37] Capasso, F., Sirtori, C., Cho, A. Coupled Quantum Well Semiconductors with Giant Electric Field Tunable Nonlinear Optical Properties in the Infrared. *IEEE J. Quantum Electron*. **30,** 1313-1326 (1994).

[38] Lee, J., Tymchenko, M., Argyropoulos, C., Chen, P. Y., Lu, F., Demmerle, F., ... & Belkin, M. A. Giant nonlinear response from plasmonic metasurfaces coupled to intersubband transitions. *Nature*, *511*(7507), 65-69 (2014).

[39] Lee, J., Nookala, N., Gomez-Diaz, J. S., Tymchenko, M., Demmerle, F., Boehm, G., ... & Belkin, M. A. Ultrathin second-harmonic metasurfaces with record-high nonlinear optical response. *Adv. Opt. Mater*, *4*(5), 664-670 (2016).

[40] Wolf, O., Campione, S., Benz, A., Ravikumar, A. P., Liu, S., Luk, T. S., ... & Brener, I. Phased-array sources based on nonlinear metamaterial nanocavities. *Nature communications*, *6*(1), 7667 (2015).

[41] Mann, S. A., Nookala, N., Johnson, S. C., Cotrufo, M., Mekawy, A., Klem, J. F., ... & Belkin, M. A. Ultrafast optical switching and power limiting in intersubband polaritonic metasurfaces. *Optica*, *8*(5), 606-613 (2021).

[42] Jeannin, M., Cosentino, E., Pirotta, S., Malerba, M., Biasiol, G., Manceau, J. M., & Colombelli, R. Low intensity saturation of an ISB transition by a mid-IR quantum cascade laser. *Applied Physics Letters*, 122(24) (2023).

[43] Cotrufo, M., Krakofsky, J. H., Mann, S. A., Böhm, G., Belkin, M. A., & Alù, A. Intersubband polaritonic metasurfaces for high-contrast ultra-fast power limiting and optical switching. *npj Nanophotonics*, *1*(1), 14 (2024).

[44] Mohajan, J., Chao, P., Jin, W., Molesky, S., & Rodriguez, A. W. Fundamental limits on radiative χ (2) second harmonic generation. Optics Express, 31(26), 44212-44223 (2023).

[45] Molesky, S., Chao, P., Jin, W., & Rodriguez, A. W. Global T operator bounds on electromagnetic scattering: Upper bounds on far-field cross sections. *Physical Review Research*, *2*(3), 033172 (2020).

[46] Miller, O. D., Polimeridis, A. G., Homer Reid, M. T., Hsu, C. W., DeLacy, B. G., Joannopoulos, J. D., ... & Johnson, S. G. Fundamental limits to optical response in absorptive systems. *Optics express*, *24*(4), 3329-3364 (2016).

[47] Lalau-Keraly, C. M., Bhargava, S., Miller, O. D., & Yablonovitch, E. Adjoint shape optimization applied to electromagnetic design. *Optics express*, *21*(18), 21693-21701 (2013).

[48] Gomez-Diaz, J. S., Tymchenko, M., Lee, J., Belkin, M. A. Nonlinear processes in multi-quantum-well plasmonic metasurfaces : Electromagnetic response, saturation effects, limits, and potentials. *Physical Review B*, *92*, 125429 (2015).

**FIGURES**

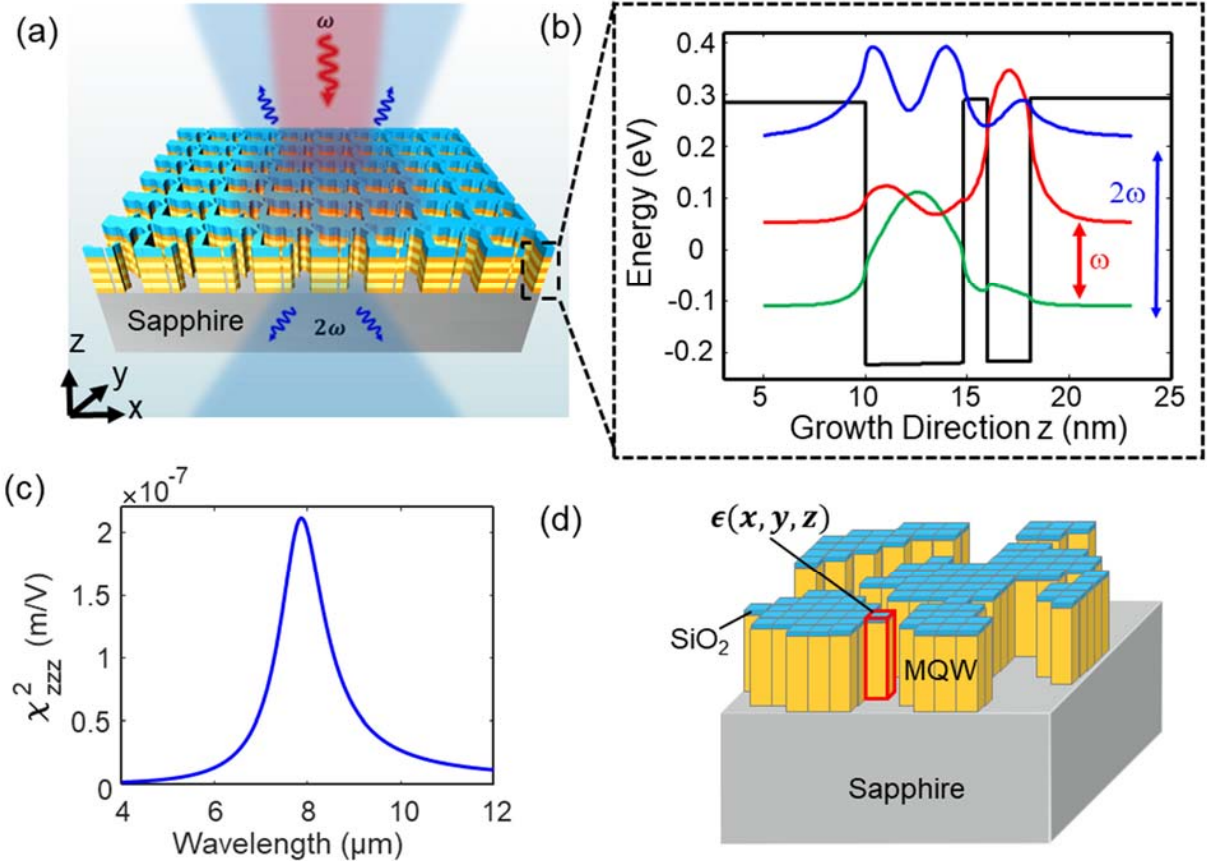

**Figure 1. Design principle of the nonlinear metasurface & schematic of the topology optimization approach.** **(a)** Schematic of the all-dielectric polaritonic nonlinear metasurface that is inverse designed using topology optimization. The individual resonators of the metasurface are made out of a nonlinear medium comprised of a multi-QW heterostructure and resides on a lower refractive index sapphire substrate. The height of all the resonators is 1.6 μm with periodicity of 4.2 μm. **(b)** 8 band k.p band structure calculation of the conduction band of a single period of the $Al_{0.52}In_{0.48}As/In_{0.53}Ga_{0.47}As$ multi-QW heterostructure used in this study. The thicknesses of the two asymmetric coupled $In_{0.53}Ga_{0.47}As$ quantum wells are optimized to support three equally space energy levels separated by 156 meV (7.9 μm) such that it can generate SH signal at 312 meV (3.95 μm). The layer sequence and thickness of the single period of the heterostructure is **20**/5/**1.2**/2.2/**20** nm, where $Al_{0.52}In_{0.48}As$ barrier layers are shown in bold. The $In_{0.53}Ga_{0.47}As$ layers are uniformly n-doped with doping density of $2\times10^{18}$ cm$^{-3}$ **(c)** Calculated value of the nonlinear susceptibility of the multi-QW structure shown in (b) as a function of the pump wavelength. **(d)** Schematic showing how topology optimization is implemented for the inverse design of the nonlinear metasurface. For optimization, the density of the material at each voxel (shown by the red cuboid) in the design domain is considered as the degree of freedom.

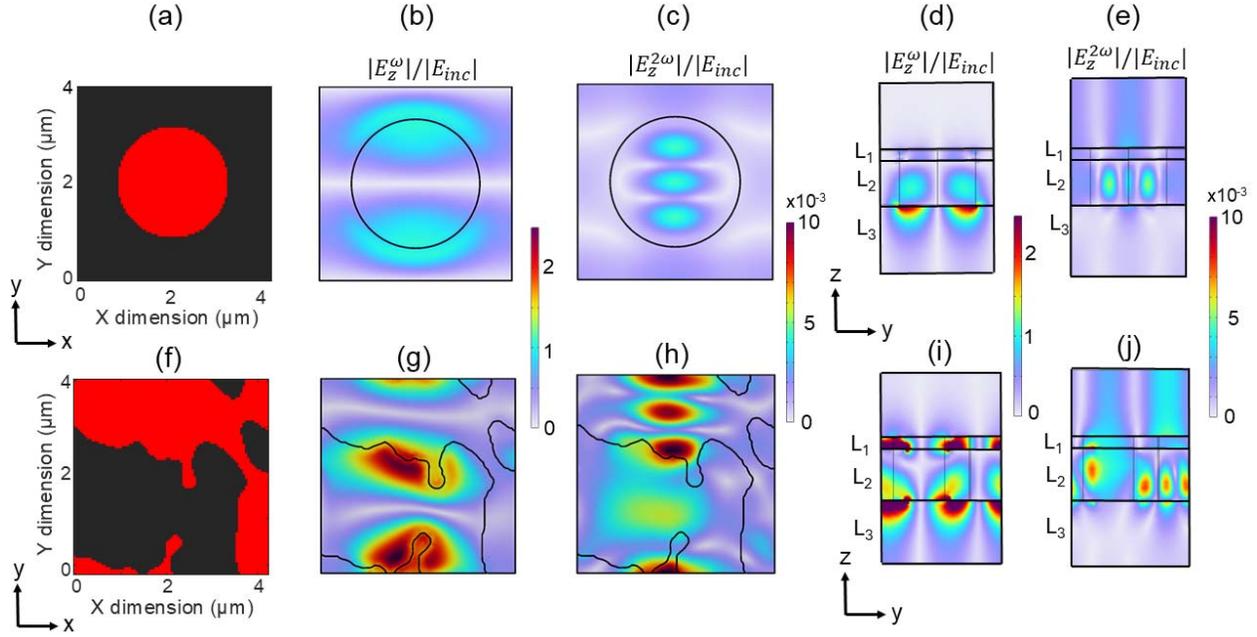

**Figure 2. Numerical simulations of nonlinear metasurfaces with inverse-designed and cylindrical resonators.** **(a, f)** In-plane (*x-y*) spatial profile of one period of the metasurfaces with cylindrical (a) and inverse-designed resonators (f). The red region corresponds to multi-QW and black region corresponds to free space. The metasurfaces are assumed to be an infinite periodic array of the resonators shown in (a) and (f). The metasurface with the cylindrical resonators is considered as a benchmark for comparing the performance of the inverse-designed metasurface. **(b, g)** Numerically calculated absolute value of the *z*-component of the electric field in the central *x-y* plane of the resonator normalized with respect to incident field $E_{inc}$ at the pump wavelength of 7.9 μm for the metasurfaces with cylindrical (b) and inverse-designed (g) resonators. **(c, h)** Numerically calculated absolute value of the *z*-component of the electric field in the central *x-y* plane of the resonator normalized with respect to incident field $E_{inc}$ at the second harmonic wavelength of 3.95 μm for the metasurfaces with cylindrical (c) and inverse-designed (h) resonators. **(d, i)** Numerically calculated absolute value of the *z*-component of the electric field in the central *z-y* plane at the center of the unit cell normalized with respect to incident field $E_{inc}$ at the pump wavelength of 7.9 μm for the metasurfaces with cylindrical (d) and inverse-designed (i) resonators. The $L_1$, $L_2$, and $L_3$ regions marked in the figure correspond to the $SiO_2$, multi-QW, and Sapphire substrate regions respectively. The fields in the multi-QW region are significantly larger for the inverse-designed metasurface. **(e, j)** Numerically calculated absolute value of the *z*-component of the electric field in the *z-y* plane at the center of the unit cell normalized with respect to incident field $E_{inc}$ at the second harmonic wavelength of 3.95 μm for the metasurfaces with cylindrical (e) and inverse-designed (j) resonators. The $L_1$, $L_2$, and $L_3$ regions marked in the figure correspond to the $SiO_2$, multi-QW, and Sapphire substrate regions respectively. The fields in the multi-QW region are significantly larger for the inverse-designed metasurface compared to the cylindrical metasurface.

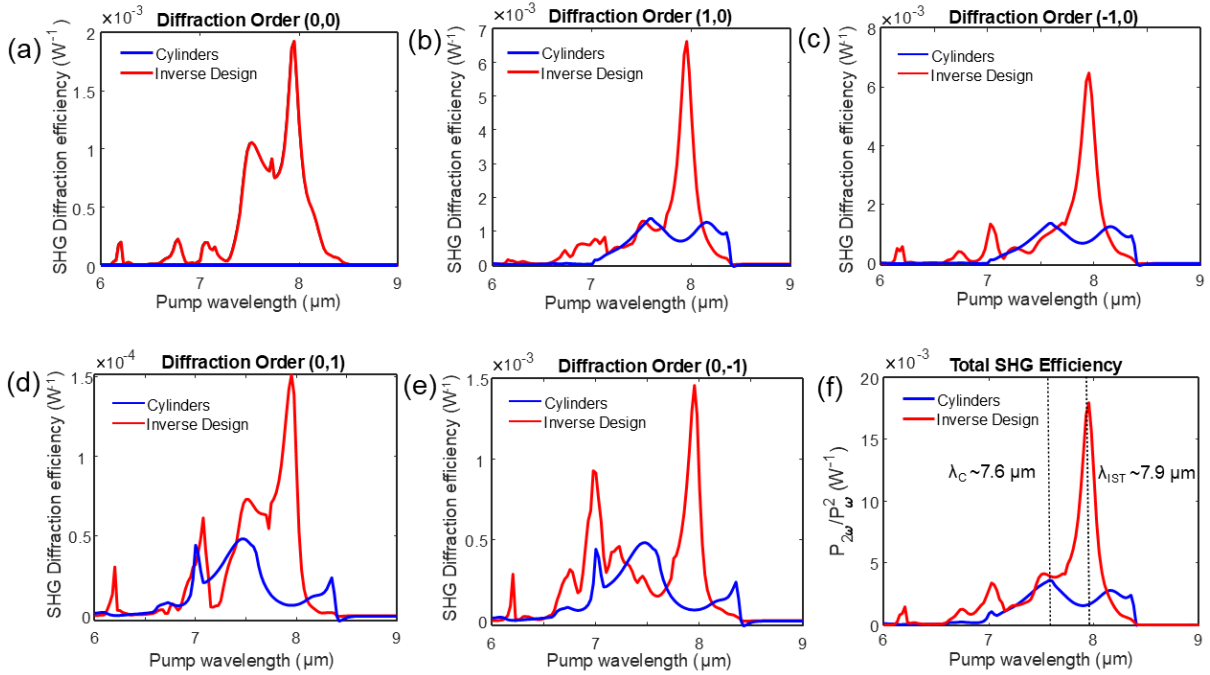

**Figure 3. Numerical simulations of the efficiency of SHG for the SH signal emitted in different diffraction orders from nonlinear metasurfaces with inverse-designed and cylindrical resonators. (a)** Calculated efficiency of SHG for the SH signal emitted in zeroth diffraction order (0,0) direction in reflection from metasurfaces with cylindrical (blue line) and inverse-designed resonators (red line). All calculations were done for a plane wave pump source incident normally on the metasurface polarized along x-direction. The metasurface with cylindrical resonators have zero emission along the normal direction because of symmetry constraints. **(b-e)** Calculated SHG intensity for the SH signal emitted from metasurfaces in different first-order diffraction directions in reflection with cylindrical (blue line) and inverse-designed resonators (red line). All calculations were done for a plane wave pump source incident normally on the metasurface polarized along x-direction. The inverse-designed metasurface has higher SHG efficiency for all diffraction directions compared to the metasurface with cylindrical resonators. **(f)** Comparison of total SHG efficiency (sum of SH emission in all diffraction orders) of the metasurface with cylindrical (blue line) and inverse-designed (red line) resonators. The metasurface with cylindrical resonators has a peak of SHG efficiency at pump wavelength $\lambda_C$ which is ~ 5x smaller compared to the peak of the SHG efficiency of the inverse-designed metasurface at pump wavelength $\lambda_{IST}$ ~ 7.9 μm.

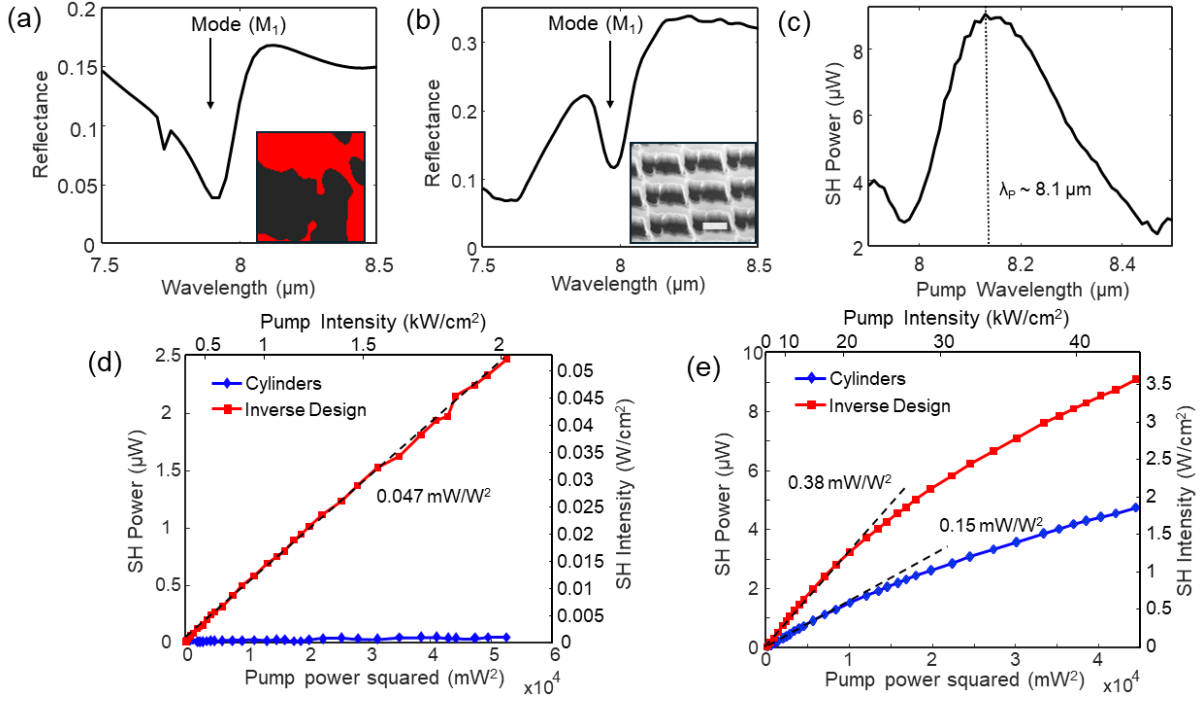

**Figure 4. Experimental characterization of the inverse-designed metasurfaces. (a)** Numerically calculated linear reflection spectrum of the inverse-designed metasurface with the unit cell shown as inset. The inverse-designed metasurface supports a mode $M_1$ that spectrally overlaps with the IST. **(b)** Experimentally measured linear reflection spectrum of the inverse-designed metasurface. The inset shows a scanning electron micrograph of the fabricated metasurface. The white scale bar corresponds to 2 μm. The measured reflection spectrum shows the dip corresponding to mode $M_1$ at the designed wavelength. **(c)** Experimentally measured SH power at normal incidence and in a reflection geometry, as a function of pump wavelength from the inverse-designed metasurface. The maximum SHG efficiency is observed at $\lambda_p \sim 8.1$ μm which is similar to the benchmark metasurface with cylindrical resonators (radius = 1.2 μm). **(d, e)** Experimentally measured SH power at pump wavelength, $\lambda_p = 8.1$ μm, measured as function of square of incident pump power for the inverse-designed metasurface (red) and the metasurface with cylindrical resonators (blue) using SH collection optics with NA = 0.18 (d) and NA = 0.85 (e) respectively. NA = 0.18 (d) captures only the SH signal emitted along the zero order where the collection optics with NA = 0.85 captures SH signal emitted along both zero and first diffraction orders. The slope of the linear fit (black dashed line) determines the nonlinear conversion factor and the values are stated in the figures.

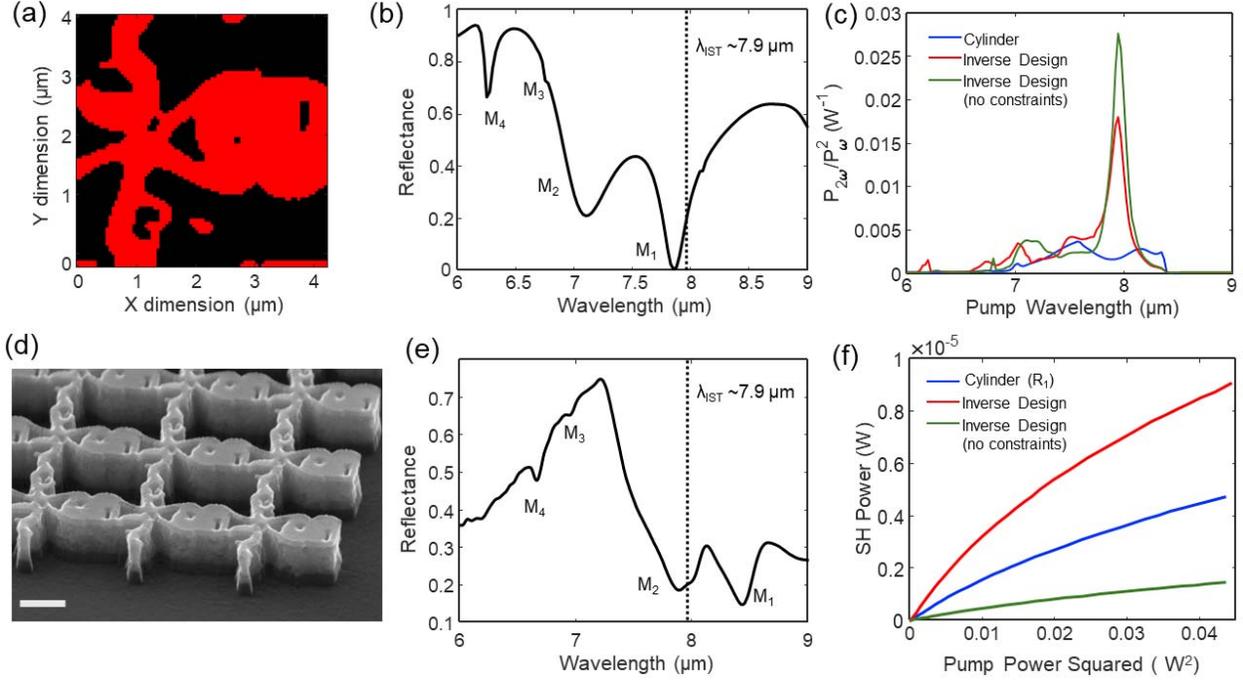

**Figure 5. Numerical simulation and experimental characterization of the metasurface designed by TO without imposing fabrication constraints. (a)** In-plane (*x-y*) spatial profile of one period of the metasurface designed by the TO algorithm without any fabrication constraints. The red region corresponds to multi-QW and black region corresponds to free space. **(b)** Numerically calculated linear reflection spectrum of the inverse-designed metasurface with no constraints. The metasurface supports 4 modes ($M_1$, $M_2$, $M_3$, and $M_4$) within the spectral region of 6-9 µm and mode $M_1$ spectrally overlaps with the state 1 to state 2 of the multi-QW IST. **(c)** Comparison of total SHG efficiency (sum of all diffraction orders) of the metasurface with cylindrical (blue line) resonators, inverse-designed metasurface with fabrication constraints (red line), and inverse designed metasurface without any fabrication constraints (green line). The metasurface with no constraints has a peak of SHG efficiency that is approximately 17 times larger compared to the peak of the SHG efficiency of the metasurface with cylindrical resonators at pump wavelength $\lambda_{IST}$ ~ 7.9 µm. **(d)** Side view scanning electron micrograph of the fabricated metasurface with resonators designed by the TO algorithm with no fabrication constraints. The white scale bar corresponds to 2 µm. **(e)** Experimentally measured linear reflection spectrum of the inverse-designed metasurface shown in (d). While all modes ($M_1$, $M_2$, $M_3$, and $M_4$) are present, they are all red shifted and mode $M_2$ spectrally overlaps with the IST resonance. **(f)** Experimentally measured SH power at pump wavelength, $\lambda_p$ = 8.1 µm, measured as function of square of incident pump power for the metasurface that was inverse-designed with constraints (red), metasurface with cylindrical resonators (radius, $R_1$ = 1.2 µm (blue)), and metasurface inverse-designed with no constraints (shown in (d)). The metasurface inverse-designed with no constraints shows the lowest SHG efficiency.